# A Refractometer for tracking the changes in the refractive index of air near 780 nm.

N. Khélifa, H. Fang, J. Xu[+], P. Juncar and M. Himbert[++]

BNM-INM; Conservatoire National des Arts et Métiers. 292, rue Saint -Martin, 75003-Paris, France.

(+) Permanent address: National Institute of Metrology, n° 7 District 11, 100013 Beijing, P.R. China.

(++) also with Laboratoire de Physique des Lasers, URA CNRS # 282, Université Paris-Nord 93430 Villetaneuse, France.

## Abstract

A new system, consisting of a double channel Fabry Perot etalon and laser diodes emitting around 780 nm is described and proposed to be used for air refractive index measurements. The principle of this refractometer is based on frequency measurements between optical laser sources. It permits quasi-instantaneous measurement with a resolution better than $10^{-9}$ and uncertainty in the $10^{-8}$ range. Some preliminary results on the stability of this system and the measurements of the refractive index of air with this apparatus are presented. The first measurements of the index of air at 780 nm are, within an experimental uncertainty of the order of $2 \times 10^{-8}$, in agreement with the predicted values by the so-called revised Edlén equations. This result is to the best of our knowledge the first to extend to the near infra-red the validity of the revised Edlén equation derived for the wavelength range 350- 650 nm.

## 1. Introduction

One of the most important sources of error limiting the accuracy of length measurements by interferometric techniques arises from the uncertainty and fluctuations of the refractive index of air. Most such interferometric techniques use, as reference sources, lasers which are frequency stabilised to atomic or molecular lines. Despite the high degree of stability (relative uncertainty below $10^{-11}$) of such reference lasers, because of the definition of the Mètre[1], the associated wavelength ($\lambda_v$) is well defined only in vacuum. Practical measurements of length or distance on the other hand are performed mostly in an air environment whose refractive index fluctuations result in variations of the wavelength ($\lambda_a = \lambda_v/n$) of the reference laser source. Generally the refractometers used for air refractive index measurements are based on the classical technique of counting interference fringes and interferometric phase measurements. In such systems the optical path of a cell of length $\ell$, placed in the interferometer, is changed by the quantity $2\ell(n-1)$ when the air is evacuated from the cell. This variation is related to the wavelength $\lambda_v$ of the source which illuminates the interferometer by the relation: $2\ell(n-1) = (k+\varepsilon)\lambda_v$ where k is an integer and $\varepsilon$ a fractional number. With this method the index of air can be measured with a level of uncertainty fixed essentially by those of $\ell$ and $\varepsilon$ (generally the uncertainty of $\lambda_v$ can be neglected).

Up to now, the best refractometers used for air refractive index measurements reach relative uncertainties limited to the $10^{-7}$ level[2]. Furthermore, their response times are rather slow compared with the rate of variation of the index of air associated with the changes of the climatic conditions (temperature, pressure, humidity and $CO_2$ concentration...). In the case of our refractometer the index of air is only determined from beat frequency measurements between optical laser sources. The uncertainties of the measurements can reach the $10^{-8}$ level. This allows one to improve the accuracy of length measurement techniques based on interferometric methods in air.

As will be show in section 2, the most important application of this apparatus concerns the development of a new type of laser reference where the wavelength is stabilised, even if n varies, as opposed to the frequency as in classical optical frequency standards[1]. This type of reference is insensitive to air refractive index fluctuations and exhibits a long term stability $\delta\lambda/\lambda$ of $10^{-8}$.

## 2. Principle of the measurement

Our apparatus is shown in figure 1. The central element of the air refractometer is a dual channel plane-plane Fabry Perot interferometer with a spacer of lengthSYMBOL $\ell = 250 mm$, made from zerodur (thermal coefficient expansion of $10^{-8}\,°C^{-1}$). This cavity will be described in more detail in section 3. Each channel is illuminated independently by a single mode laser diode operating around 780 nm. One channel can be evacuated when the other is kept at atmospheric pressure. The beat frequency between the laser diodes LD1 and LD2 is measured using the photo-detector PD3 and that between LD1 and LD3 by PD4. The photodiodes PD1 and PD2 are used to lock the lasers LD1 and LD2 to transmission peaks of their respective cavities. When the frequency of a laser is locked to a transmission peak of this Fabry Perot etalon, it follows that $\ell = k(\lambda/2)$, where k is an integer and $\lambda$ the wavelength of the laser.

The method of the measurement is as follow.

In a first step, air is present in both channels of the interferometer. With the help of a beat frequency measurement, using the photo-detector PD3, it is straightforward to ensure that each laser is locked to the same peak k of frequency $\nu_k = k\dfrac{c}{2n\ell}$. ( $c$ is the speed of light in vacuum and $n$ is the refractive index of air). In this situation, $\Delta\nu = \nu_1 - \nu_2$ is in principle equal to zero and since the wavelengths $\lambda_{1a}$ and $\lambda_{2a}$ (both in air) are equal to $\lambda_{1a} = \lambda_{2a} = 2\ell/k$, the corresponding frequencies are given by:

$$\nu_1 = \nu_2 = \frac{c}{n\lambda_{1a}} = \frac{c}{n\lambda_{2a}}. \tag{1}$$

When the pressure of the first channel, illuminated by the laser LD1, is slowly reduced, the laser LD1 remains locked to the same peak k. After evacuation of this channel (the residual pressure is about 1 Pa), the wavelength $\lambda_{1v}$ in vacuum remains equal to:

$$\lambda_{1v} = 2\ell/k = \lambda_{2a}. \tag{2}$$

This means that the interference pattern between the two mirrors of the Fabry Perot cavities remains unchanged during the pumping procedure. On the other hand the frequency of the laser LD1 becomes:

$$\nu_1^* = c/\lambda_{1v} \tag{3}$$

giving rise to a beat frequency $\Delta\nu = \nu_1^* - \nu_1$ between the two diode lasers equal to $(n-1)\nu_1$ which is of order 105 GHz. The new frequency $\nu_1^*$ of the laser LD1 can be calibrated by comparison with a reference frequency $\nu_{ref}$, in our case a laser LD3 locked to a hyperfine component of the rubidium $D_2$ line, Thus:

$$\nu_1 = \nu_1^* - \Delta\nu \quad \text{and} \quad \nu_1^* = \nu_{ref} + \Delta\nu^* \tag{4}$$

Finally, from our knowledge of $\nu_{ref}$ and the almost instantaneous measurements of $\Delta\nu$ and $\Delta\nu^*$, we have an instantaneous determination of the refractive index of air. Explicitly:

$$n = \frac{\nu_1^*}{\nu_1} = \frac{\nu_{ref} + \Delta\nu^*}{\nu_{ref} + \Delta\nu^* - \Delta\nu} \tag{5}$$

From (5), one can see that with this technique the measurement of the index of air does not depend on the value of k nor on $\ell$ which is generally not the case of the other traditional interferometric methods.

There are two main difficulties inherent to our method. The first is to scan continuously the frequency of the laser LD1 over about 105 GHz while it remains locked to the same transmission peak of the F-P without mode hopping. The second is to measure the associated beat frequency $\Delta\nu$ which lies close to this value. To address the first problem we employ distributed Bragg reflector (DBR) laser diodes (Yokogawa company). As well as providing single mode operation with a narrow spectral linewidth (1 MHz), these diodes can be tuned continuously over a wide wavelength range ( >150 GHz ). The second difficulty, namely the large frequency difference measurement can be overcomes by using two possibilities:

- The first is a direct measurement of $\Delta\nu$ and use three wave mixing on a Schottky diode[3]. Here the 2$^{nd}$ harmonic of a stabilised microwave field of frequency $\Omega_{RF} \approx 52$ GHz with the optical laser beams of frequencies $\nu_1$ and $\nu_1^*$. This give a beat signal at frequency $\left((\nu_1^* - \nu_1) - 2.\Omega_{RF}\right)$ which can be measured directly.

- The other method use the fact that the beat frequency $\Delta\nu$ can be compared to the free spectral range (FSR=$c/2n\ell$) of the Fabry Perot. In this case we measure the frequency difference between $\nu_1^*$ and $\nu_1 + p\dfrac{c}{2n\ell}$ . The integer p is such that this difference is less than 1 FSR. More details about this procedure are provided in section 5.

In practice, $\nu_{ref}$ is known to within a few kilohertz and $\Delta\nu$ and $\Delta\nu^*$ can be measured with an accuracy well below 1 MHz. This means that we can determine n with a relative uncertainty better than $10^{-8}$ (one standard deviation). An interesting characteristic of this refractometer is the fact that the value of the wavelength $\lambda_{2a}$ of the laser LD2, is insensitive to air refractive index fluctuations. In fact, from the relations (2), (3) and (4) the wavelength of the laser LD2 is given by the expression: $\lambda_{2a} = c/(\nu_{ref} + \Delta\nu^*)$.

This property is used at present in our laboratory to provide an optical wavelength reference.

## 3. Construction and design of the refractometer

In order to minimise the deformations of the mirrors when making the vacuum inside the channel 1, the double Fabry Perot etalon is made from a single piece of zerodur on which two fused silica 20 mm thick mirrors, with appropriate coatings, are fixed by optical contact (figure 2). Behind each mirror lies a 5 mm thick silica spacer (of the same geometry as the zerodur spacer). One side of this spacer is fixed to the mirror by optical contact. The other side, supports a 20 mm thick optically flat silica window also fixed by optical contact. As one can see in figure 2, there are holes in the mirrors in order to equalise the pressure on each side of the mirror during the pumping procedure. With this geometry, when the central cavity of the Fabry Perot is evacuated, the small deformations undergone by the windows will not affect the flatness of the mirrors.

## 4. Preliminary tests

### 4.1- Frequency stability measurements

In order to characterise the stability of the interferometer the fluctuations $\delta v_i$ of the frequency $v_k = k\frac{c}{2\ell}$ of the Fabry Perot are measured by comparing $v_k$ with $v_{ref}$ using a heterodyne method (figure 3). The reference frequency $v_{ref}$ is obtained using saturated absorption spectroscopy of atomic rubidium (Rb) in a vapour cell. We employ the crossover resonance between the hyperfine transitions $\left(5S_{1/2}, F_g = 2 \to 5P_{3/2}, F_e=2\right)$ and $\left(5S_{1/2}, F_g = 2 \to 5P_{3/2}, F_e= 3\right)$ of the $^{87}$Rb-D$_2$ line. The frequency of this resonance is given by [4]: $v_{ref} = 384\,227\,981,877$ MHz

The beat frequency $\delta v_i$, between this reference and the laser diode LD1 locked to the transmission peak of frequency $v_k$ of one cavity of the Fabry Perot, is measured during an integration time $\tau$ of one second. N equivalent measurements made at equal time intervals $\tau$ allow one to perform a statistical analysis of the frequency fluctuations $\delta v_i$ via the relative Allan variance[5]:

$$\sigma = \frac{1}{v_k}\left(\sum_{i=1}^{N}\frac{(\delta v_i - \delta v_{i-1})^2}{2(N-1)}\right)^{1/2}$$

where $\delta v_i = v_k(t_i) - v_{ref}$ with $t_i = i.\tau$ and $i = 1,2,...N$.

The results presented in figure 4 show the good short term stability of the Fabry Perot ($\approx 5 \times 10^{-11}$ in 100 s), which corresponds to residual frequency fluctuations. Figure 4 shows also the stability of the reference frequency, measured by the same technique using a second system ($\approx 4 \times 10^{-12}$ in 100 s).

The long term stability of the interferometer is limited by the temperature changes of the zerodur. The linear thermal expansion coefficient of this material has been evaluated by measuring, over a long period of time, the shift in the beat note between our reference and a laser diode locked on the peak of frequency $v_k$ of the Fabry Perot. Figure 5 shows a decrease of $v_k$ by about 18 MHz $°C^{-1}$, which corresponds to a relative increase in the length of the interferometer by about $5 \times 10^{-8}$ $°C^{-1}$. This value agrees with the longitudinal thermal expansion coefficient in the range of $0\text{-}50\,^0C$, of class I zerodur glass ceramics ($\pm 5 \times 10^{-8}$ $°C^{-1}$) specified by the company Schott. In actual fact, since thermal fluctuations lead to almost identical changes in the lengths of both cavities, their influence upon our measurements can be neglected.

**4.2- Effect of parallelism of mirror faces and air pressure**

In order to measure the refractive index of air with a relative uncertainty in the range of $10^{-8}$ we require that the length difference between the two cavities of the Fabry Perot (central and peripheral cavity) remain the same to within $\lambda/300$. This corresponds to a shift of the beat note between the corresponding peak k of the two cavities smaller than 4 MHz. Such a shift can be measured easily. Any variation of the length of the cavities due to a pressure differential can be measured by the following procedure which we now describe.

Initially the two cavities are in air and we choose a peak number k of the peripheral cavity such that the corresponding frequency $v_2$ is close to the frequency $v_{ref}$. The quantity $\Delta v = v_2 - v_{ref}$ is measured. When the central cavity is evacuated, the external pressure leads to a reduction in the length giving rise to a frequency shift of the peak k:

$$v_2^j = v_2 + c_2.$$

The term $c_2$ is of order 30 MHz corresponding to a change of the cavity length by $\approx \lambda/40$. To first order, the corresponding deformation is assumed to be equivalent in both cavities. Since our method of measurement is independent of the length of the F-P cavities, linear changes of this length will not affect the accuracy of the measurements. However if we place an upper limit for the residual non-linearity of the deformation between the two cavities of 10%, this will introduce a differential frequency shift of order 3 MHz ($\lambda/400$) which corresponds to a relative uncertainty of order $8 \times 10^{-9}$ in the measurement of n. This effect is in fact the principal limitation on the accuracy of our apparatus.

### 4.3- Sensitivity of the apparatus to misalignment errors

A small change $\delta\theta$ in the inclination of the laser beam with respect to the optic axis of the Fabry Perot interferometer induces a corresponding modifications of the path difference. This introduces, first a deformation of the shape of the transmission peak (this deformation can easily be seen when the interferometer is illuminated with a periodic laser frequency ramp) and second, a variation of the beat frequency between the reference frequency and the laser locked to the Fabry Perot. A relative uncertainty of order $10^{-8}$ needs angular adjustment to within $10^{-4}$ rad.

This implies a beat frequency measurement uncertainty below 4 MHz. This requirement is easily satisfied by observing the position of the beat frequency signal on the spectrum analyser. Consequently we can neglect errors due to misalignments.

## 5. Determination of the refractive index of air: first measurements

Since the Schottky diode and the RF mixer system is not yet operational the beat frequency ($\Delta \nu \approx 105$ GHz) is measured indirectly. Our method involves a similar concept which is used to achieve absolute distance measurement[6]. The results obtained for the refractive index of air are compared with those derived from the revised Edlén equation[7] used for ambient atmospheric conditions over the range of wavelengths 350- 650 nm.

### 5.1- Measurement procedure

First, we start with both cavities of the Fabry Perot at atmospheric pressure. The frequency $\nu_k$ of the transmission peak, on which the diode laser LD1 (for central cavity) and LD2 (for peripheral cavity) are locked, is respectively $\nu_1$ and $\nu_2 = \nu_1 + c_{12}$. The correction term $c_{12}$ is of order 70 MHz corresponding to the shift between the peak k of the two cavities. When the central cavity of the interferometer is slowly evacuated the frequency $\nu_1$ changes and goes to $\nu_1^*$. This last frequency is obtained by measuring the beat frequency between $\nu_1^*$ and $\nu_r$. For more clarity in the description of the measurement procedure the different frequencies, used here, are indicated in figure 6.

The frequency $\nu_1$ is chosen such that the difference between $\nu_1^*$ and $\nu_r$ corresponds to a few hundred megahertz in order that it can be measured by conventional beat frequency techniques:

$$\nu_1^* = \nu_r + \delta\nu_{1r}. \tag{6}$$

The refractive index of air is given by the ratio $\dfrac{v_1^*}{v_1^j}$, where $v_1^j$ is the frequency $v_1$ shifted by $c_1$ due to the fact that the central cavity is under vacuum.

$$v_1^j = v_1 + c_1 \tag{7}$$

During the evacuation of the central cavity of the interferometer the peripheral channel is maintained at atmospheric pressure but undergoes a similar compression effect giving a frequency $v_2^j$ related to $v_2$ by a correction term $c_2$

$$v_2^j = v_2 + c_2 . \tag{8}$$

Next, the laser LD2 is no longer locked to the $k^{th}$ peak of frequency $v_2$ but instead is scanned manually and, after we have counted an integer number p of fringes, it is relocked to the peak number $(k+p)$ such that the frequency of this laser is varied from $v_2^j$ to $v_2^*$.

$$v_2^* = v_2^j + p\dfrac{c}{2n\ell} \tag{9}$$

The frequency $v_2^*$ lies close to $v_1^*$ to within less than one FSR of the Fabry Perot, i.e. $\Delta v_{FP}^{air} = \dfrac{c}{2n\ell}$. The calibration of $v_2^*$ in terms of $v_r$ gives:

$$v_2^* = v_r + \delta v_{2r} \tag{10}$$

where $\delta v_{2r}$ represents the beat frequency between the reference source and the diode laser LD2 at frequency $v_2^*$. From the relations (6), (7), (8), (9) and (10) we obtain:

$$v_1^j = v_2^j - c_2 - c_{12} + c_1 \quad \text{and} \quad v_2^j = v_r + \delta v_{2r} - p\dfrac{c}{2n\ell} . \tag{11}$$

Finally, from (11) and (6) the refractive index of air can be expressed in the form:

$$n = \frac{v_r + \delta v_{1r} + p\frac{c}{2\ell}}{v_r + \delta v_{2r} + c_1 - c_2 - c_{12}} \quad . \tag{12}$$

To first order, we suppose that the corrections $c_1$ and $c_2$ are the same which means that the atmospheric pressure causes the lengths of both cavities to change by the same amount. With this procedure, we can make measurements of the refractive index of air and also track its fluctuations at any time. This is achieved thanks to the quasi-instantaneous measurement of the two beat frequencies $\delta v_{1r}$ and $\delta v_{2r}$.

With this method, the uncertainty in the index of air is less than $2 \times 10^{-8}$. This limitation is mainly due to the uncertainty in the correction $(c_1 - c_2)$. In our case, all the corrections have been evaluated to within 3 MHz (one standard deviation) while the FSR in vacuum was measured in our laboratory, by a beat frequency technique to be $599{,}594 \pm 0{,}005$ MHz. Figure 7 shows the results of the measurement carried out continuously during a few hours for different days. The calculated values are obtained from the measured values of temperature, atmospheric pressure and air humidity by using the revised Edlén equations[7]:

$$(n-1)_{tp} = \frac{(p/Pa)(n-1)_s}{96095{,}43} \times \frac{\left[1 + 10^{-8}(0{,}601 - 0{,}00972\ t/°C)p/Pa\right]}{(1 + 0{,}0036610\ t/°C)}$$

where $(n-1)_{tp}$ is the refractive index of air at a temperature t and pressure p and $(n-1)_s$ is given by the dispersion equation :

$$(n-1)_s \times 10^8 = 8342{,}54 + 2406147\left[130 - (\lambda^{-1}/\mu m^{-1})^2\right]^{-1} + 15998\left[38{,}9 - (\lambda^{-1}/\mu m^{-1})^2\right]^{-1}$$

where $\lambda^{-1} = 1/\lambda$ represent the wavenumber expressed in $\mu m^{-1}$.

For moist air containing a partial pressure f of water vapour the refractive index is:

$$n_{tpf} = n_{tp} - (f/Pa) \times \left[3{,}7345 - 0{,}0401(\lambda^{-1}/\mu m^{-1})^2\right] \times 10^{-10}$$

The uncertainty in the calculated values, about $2 \times 10^{-7}$, is associated with the resolution of the instruments which are used for determining air temperature, pressure and humidity. The absolute values of the refractive index measured by beat frequency technique are in same range as those calculated above, but more accurate.

## 6. Conclusion

We have developed a novel refractometer based on a two channel Fabry Perot interferometer illuminated by tuneable frequency laser diodes. Air refractive index measurements are made simply by a beat frequency technique which provides a resolution better than $10^{-9}$. The first results with this apparatus are in agreement with those calculated with the empirical Edlén Formulae which are commonly used for calculation of the index of air but the relative uncertainty of our measurements reach the level $2 \times 10^{-8}$. We can remark that the revised Edlén formulae derived for the wavelength range of 350- 650 nm are also valid near 780 nm to within $10^{-7}$.

This work is supported by the Bureau National de Métrologie and by the Conseil Régional d'Ile de France (LUMINA contract # E- 946-95/98). We thank M.D. Plimmer for critical reading of the manuscript

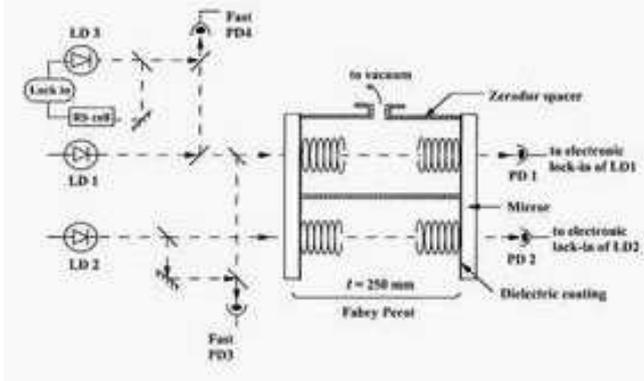

Figure 1

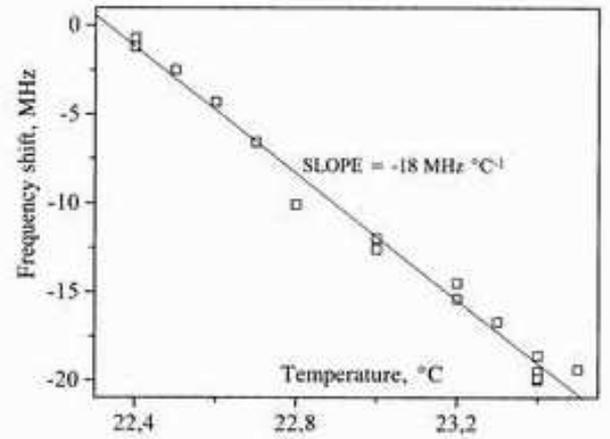

Figure 5

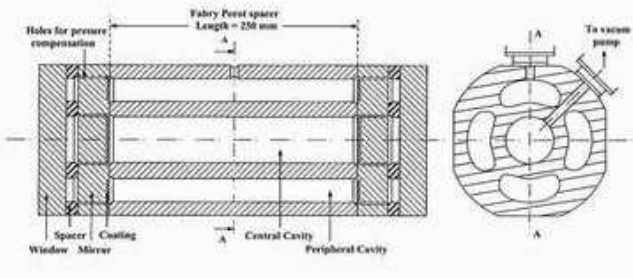

Figure 2

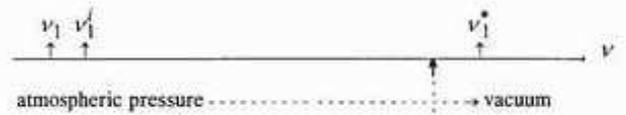
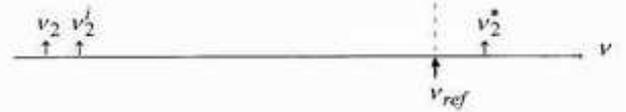

Figure 6

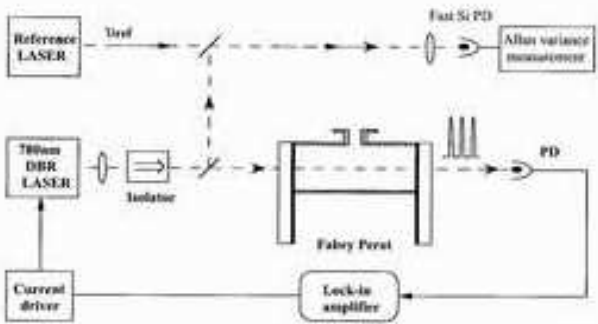

Figure 3

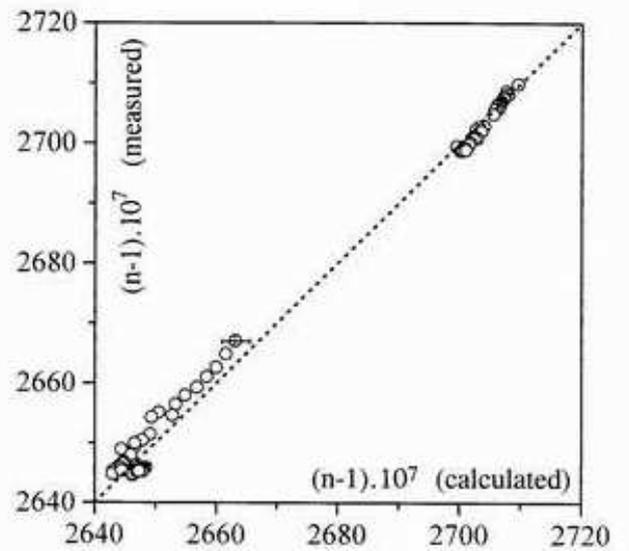

Figure 7

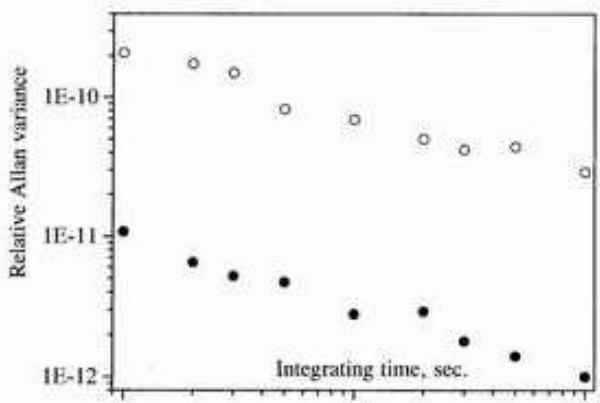

Figure 4

# Figure Captions

**Figure 1**: Experimental set-up of our refractometer. LD are lasers diode and PD are photodiodes.

**Figure 2**: Design (longitudinal and transverse sections) of the Fabry Perot etalon.

**Figure 3**: Block diagram for stability measurements of the Fabry Perot.

**Figure 4**: Relative standard deviation plot against integration time.

( • ) stability of the laser diode locked to $D_2$ line of rubidium.

( o ) stability of the laser diode locked to a transmission peak of the Fabry Perot etalon.

**Figure 5**: Change of the beat frequency between the reference frequency and the laser diode locked to the transmission peak of one cavity of the Fabry Perot versus temperature.

**Figure 6**: A representation of the procedure used for the measurements of the refractive index of air.

**Figure 7**: Plot of measured values of the refractive index of air and those calculated by the revised Edlèn equation. The error bars are derived from the uncertainties quoted in the various optical measurements and instruments. The dashed curve represents the bisecting line.